\documentclass[
showpacs,
floatfix,
aps,
prb,
amsmath,
twocolumn,
superscriptaddress,
tightenlines
groupaddress,
]{revtex4}

\usepackage{graphicx}
\usepackage{dcolumn}
\usepackage{amsmath}
\usepackage{bm}
\usepackage{times}
\usepackage[varg]{txfonts}

\def\eac{\epsilon_{\mbox{{\scriptsize ac}}}}
\def\edc{\epsilon_{\mbox{\scriptsize dc}}}

\def\pac{\phi_{\mbox{\scriptsize ac}}}

\def\oc{\omega_{\mbox{\scriptsize {C}}}}
\def\oh{\omega_{\mbox{\scriptsize {H}}}}

\def\rc{R_{\mbox{\scriptsize {C}}}}

\begin{document}
\title{
Non-linear magnetotransport in microwave-illuminated two-dimensional electron systems
}
\author{A. T. Hatke}
\affiliation{School of Physics and Astronomy, University of Minnesota, Minneapolis, Minnesota 55455, USA} 
\author{H.-S. Chiang}
\affiliation{School of Physics and Astronomy, University of Minnesota, Minneapolis, Minnesota 55455, USA} 
\author{M.\,A. Zudov}
\email[Corresponding author: ]{zudov@physics.umn.edu}
\affiliation{School of Physics and Astronomy, University of Minnesota, Minneapolis, Minnesota 55455, USA} 
\author{L.\,N. Pfeiffer}
\author{K.\,W. West}
\affiliation{Bell Labs, Alcatel-Lucent, Murray Hill, New Jersey 07974, USA}
\received{18 April 2008}

\begin{abstract}
We study magnetoresistivity oscillations in a high-mobility two-dimensional electron system subject to both microwave and dc electric fields.
First, we observe that the oscillation amplitude is a periodic function of the inverse magnetic field and is strongly suppressed at microwave frequencies near half-integers of the cyclotron frequency. 
Second, we obtain a complete set of conditions for the differential resistivity extrema and saddle points.
These findings indicate the importance of scattering without microwave absorption and a special role played by microwave-induced scattering events antiparallel to the electric field.

\end{abstract} 
\maketitle

Over the past few years, it was realized that dc resistivity of a high-mobility two-dimensional electron system (2DES) in very high Landau levels (LLs) exhibits a variety of unexpected features when subject to external electric fields and sufficiently low temperatures.
Relevant phenomena include microwave (ac)-induced resistance oscillations (MIRO)\citep{zudov:2001a,ye:2001} and Hall field (dc)-induced resistance oscillations (HIRO).\citep{yang:2002a}
Microwave-illuminated 2DES became a subject of intense experimental\citep{miroexperiment1} and theoretical\citep{mirotheory} interest, following the discovery of zero-resistance states (ZRS).\citep{mani:2002,zudov:2003,andreev:2003}
More recently it was demonstrated that the effects of dc electric field\citep{bykov:2007,zhang:2008} (or its combination with microwaves\citep{zhang:2007c}) on electron transport can also be quite dramatic leading to zero-differential resistance states.

MIRO and HIRO are periodic in inverse magnetic field $1/B$ and originate from inter-LL transitions owing to microwave absorption and elastic scattering off short-range disorder, respectively.
MIROs are governed by a parameter $\eac\equiv\omega/\oc$, where $\omega=2\pi f$ is the microwave frequency and $\oc=eB/m^*$ is the cyclotron frequency.\citep{zudov:2001a} 
HIROs in turn are controlled by $\edc\equiv \oh/\oc$, where $\hbar\oh \simeq 2e E\rc$ ($E$ is the Hall field and $\rc$ is the Larmour radius) is the Hall voltage across the cyclotron orbit.\citep{yang:2002a}
Experimentally, MIRO maxima\citep{zudov:2004,studenikin:2005}  are found at $\eac \simeq n-\pac$ ($n, m \in \mathbb{Z}^+$, throughout this paper; $\pac\lesssim 1/4$) and HIRO maxima\citep{zhang:2007a,vavilov:2007,lei:2007a} (measured in differential resistance $r=dV/dI$) occur at $\edc \simeq n$. 

Recently, experimental studies of 2DES were extended into the regimes where MIRO and HIRO coexsist.\citep{zhang:2007c}  
Magnetic field sweeps under microwaves illumination at fixed $I$ suggested that the peaks in $r$ occur at
\begin{equation} 
\eac + \edc \simeq n.
\label{acdc}
\end{equation}
Equation (\ref{acdc}) represents an important result since it indicates a dominant role of combined inter-LL transitions that consist of an energy jump due to microwave absorption and a space jump in the direction {\em parallel} to the electric field due to scattering off impurities.\citep{miroexperiment1,zhang:2007c}
However, there are several issues that remained unresolved.
First, while Eq.\,(\ref{acdc}) correctly reproduces HIRO in the limit of vanishing $\eac$, it fails to describe MIRO.
Second, according to Eq.\,(\ref{acdc}) the resistivity peaks should {\em continuously} move to higher $B$ (lower $\eac$) with increasing dc $I$ (higher $\edc$).
Instead, it is observed that the peaks first evolve into the minima without changing their positions and then {\em abruptly} jump to catch up with Eq.\,(\ref{acdc}).
Finally, while Eq.\,(\ref{acdc}) defines the line in the $(\eac,\edc)$ plane around which the maxima of $r$ are likely to be found, the precise conditions for $\eac$ and $\edc$ remain unknown.

It was also recognized\citep{zhang:2007c} that at some special values of $\eac$, e.g. $\eac\simeq n+1/2$, dc electric field has surprisingly little effect on $r$.
This observation appears rather puzzling, since in the absence of microwave radiation dc field gives rise to a drop in $r$, which can be linked to suppression of elastic impurity scattering \citep{zhang:2007a,lei:2007a} or inelastic scattering due to dc-induced nonequilibrium distribution of electrons.\citep{zhang:2007b, vavilov:2007}
This drop can be quite significant as demonstrated by recently reported formation of dc-induced zero-differential resistance states.\cite{bykov:2007,zhang:2008}

In this paper, we address all of the above issues.
We first examine the amplitude, the period, and the phase of differential resistivity oscillations in 2DES subject to microwaves and dc electric fields both as a function of $B$ at fixed $I$ and as a function of $I$ at fixed $B$. 
We then analyze differential resistivity maps in the $(\eac,\edc)$ plane and establish the additional condition that, together with Eq.\,(\ref{acdc}), provides a complete set of equations for the resistivity maxima.
The results indicate that differential resistivity maxima occur when combined transitions parallel to the electric field are maximized while those antiparallel to the electric field are minimized.
Finally, inclusion of impurity scattering without microwave absorption allows us to explain oscillation suppression at $\eac \simeq n+1/2$ as well as deviations from Eq.\,(\ref{acdc}).

Our Hall bar sample with lithographic width $w = 100$ $\mu$m was fabricated from a symmetrically doped GaAs/Al$_{0.24}$Ga$_{0.76}$As 300-\AA-wide quantum well grown by molecular-beam epitaxy. 
Ohmic contacts were made by evaporating Au/Ge/Ni and thermal annealing in forming gas.
After brief illumination at several kelvin, the electron mobility $\mu$ and the density $n_e$ were $\simeq 1.2 \times 10^7$ cm$^2$/Vs and $3.7 \times 10^{11}$ cm$^{-2}$, respectively. 
The experiment was performed at a constant coolant temperature $T \simeq 1.5$ K in a $^3$He cryostat equipped with a superconducting solenoid. 
Microwave radiation of $f=69$ GHz was generated by a Gunn diode seeding a passive frequency doubler.
Differential resistance $r$ was measured by using quasi-dc (a few hertz) lock-in technique.

\begin{figure}[t]
\includegraphics{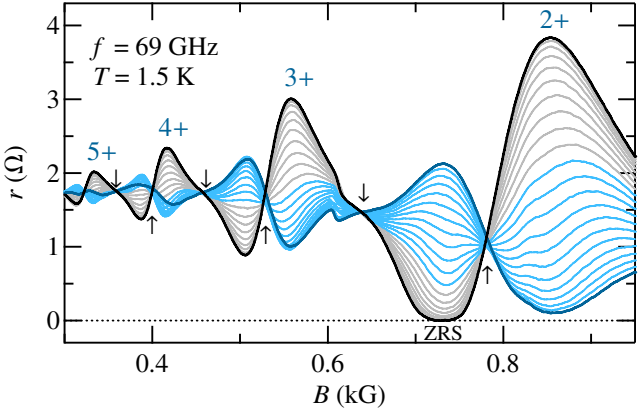}
\caption{(color online)
Microwave magnetoresistance $r$ under dc excitations, from $I = 0\,\mu$A [black curve] to $I = 22\,\mu$A [dark blue (dark grey) curve], in 2 $\mu$A increments. 
Integers show the order of the MIRO peaks.
The arrows mark zero-response nodes that largely remain immune to dc excitation.
}
\vspace{-0.15 in}
\label{acdc_tr}
\end{figure}
In Fig.\,\ref{acdc_tr}\,(a), we show the differential resistance $r$ under microwave illumination acquired by sweeping magnetic field $B$ at different constant dc $I$, from 0 to 22 $\mu$A, in 2 $\mu$A increments.
At zero dc bias [cf. black curve], we observe a sequence of MIRO maxima (marked by $n+$) and a well developed ZRS. 
With increasing $I$, maxima (minima) evolve into the minima (maxima) without obvious change in the $B$ positions expected from Eq.\,(\ref{acdc}).
Further, we observe a series of crossing points near $\eac\simeq n, n+1/2$ [cf.\,$\uparrow,\downarrow$], where $I$ has little effect on microwave photoresistance.
One can notice that for half-integer $\eac$, this behavior is considerably more robust than for integer $\eac$.

\begin{figure}[t]
\includegraphics{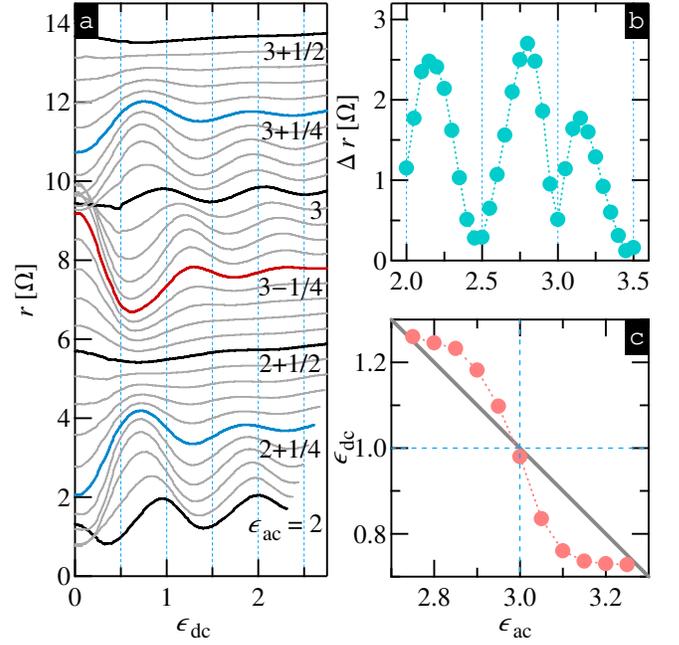}
\caption{
(a) Differential resistance $r$ vs $\edc$ at fixed $\eac$ from 2 to 3.5 in steps of 0.05.
Traces are vertically offset in increments of 0.4 $\Omega$. 
(b) Amplitude of the oscillations vs $\eac$.
(c) Position of the resistance maximum extracted from (a) near $(\eac,\edc)=(3,1)$. 
Solid line corresponds to Eq.\,(\ref{acdc}).
}
\vspace{-0.15 in}
\label{edc}
\end{figure}

In Fig.\,\ref{edc}(a), we present $r$ as a function of $\edc$ at fixed $B$ corresponding to fixed $\eac$ from $2$ to $3+1/2$ (bottom to top) in steps of $0.05$.
Traces are vertically offset for clarity in increments of 0.4 $\Omega$.
There are several issues of interest here: the amplitude of the oscillations, the period, and the phase.
We first examine the evolution of the amplitude of the oscillations with increasing $\eac$.
Even without compensating for the decay due to the Dingle factor, one clearly observes non-monotonic dependence of the amplitude on $\eac$.
The oscillations are maximized at the MIRO extrema, $\eac \simeq n \pm 1/4$, and are strongly suppressed near $\eac\simeq n, n + 1/2$.
Further examination reveals substantial difference between integer and half-integer $\eac$; while the oscillations are clearly visible at integer $\eac=2,3$, they virtually disappear at $\eac=2+1/2,3+1/2$.
In Fig.\,\ref{edc}(b), we show the amplitude of the oscillations $\Delta r$ as a function of $\eac$ and conclude that the oscillation strength is largely determined by the magnitude of the linear-response microwave photoresistance.

We now turn to the period and the phase of the oscillations.
First, we observe that the data at integer $\eac$ [cf.\,$\eac=2,3$] closely mimic HIRO\citep{zhang:2007a} with the maxima occurring at $\edc \simeq n$.
In contrast, traces corresponding to the MIRO maxima and minima [cf.\,$\eac\simeq 3-1/4$\ and $\eac\simeq 3+1/4$] are shifted by a $1/4$ cycle in the opposite directions and are out of phase.
Both of these observations are consistent with Eq.\,(\ref{acdc}).
However, at the intermediate values of $\eac$, e.g. at $3\lesssim \eac \lesssim 3+1/4$, we observe systematic deviations from Eq.\,(\ref{acdc}).
Indeed, once $\eac$ starts to increase (decrease) from $\eac=3$ the peak at $\edc \simeq 1$ moves to lower (higher) $\edc$ by $\simeq 1/4$ much faster than expected.
This is illustrated in Fig.\,\ref{edc}(c) where we plot the peak position in the $(\eac,\edc)$ plane, which is extracted from Fig.\,\ref{edc}(a) [circles], and the behavior predicted by Eq.\,(\ref{acdc}) [straight line].
It confirms that Eq.\,(\ref{acdc}) correctly describes the end and middle points of the considered interval but fails to account for the evolution at the intermediate values.

\begin{figure}[t]
\includegraphics{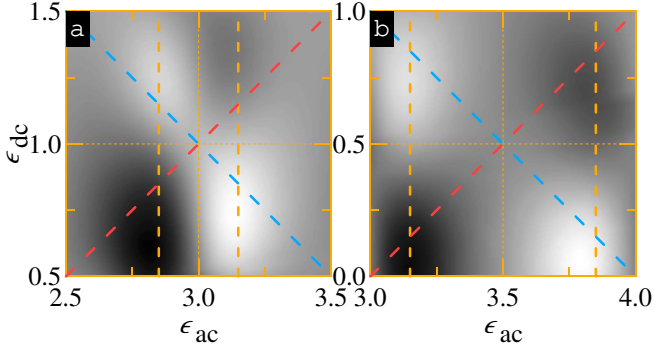}
\caption{(a)[(b)] Greyscale plot of differential resistivity $r$ in the $(\eac,\edc)$ plane around $(\eac,\edc)=(3,1)$\,[$(\eac,\edc)=(7/2,1/2)$].
Dashed lines correspond to integer values of $\eac+\edc$\,(negative slope) and $\eac-\edc$\,(positive slope).
Vertical lines drawn at the $\eac=\eac^{\pm}$ pass through the extrema in $r$.
}
\vspace{-0.15 in}
\label{image}
\end{figure}

We now examine the conditions for the differential resistivity maxima and minima.
In Fig.\,\ref{image}(a) we present greyscale intensity plot of the differential resistivity in the $(\eac,\edc)$ plane in the vicinity of $(\eac,\edc)=(3,1)$.
The data readily reveal that the maxima (light) appear next to the diagonal given by $\eac+\edc=4$, in general agreement with Eq.\,(\ref{acdc}).
We further observe that the maxima are roughly symmetric about $(\eac,\edc)=(3,1)$ and their corresponding $\eac$ and $\edc$ are given by $\eac \simeq n \mp 1/4$ and $\edc \simeq m \pm 1/4$, respectively.
Minima show similar characteristics but appear along a diagonal given by $\eac-\edc=2$.
We thus conclude that maxima$^+$ and minima$^-$ are symmetrically offset from $(\eac,\edc)=(m,n)$ by $\simeq \pm 1/4$,
\begin{align}
(\eac,\edc)^+\simeq(n\pm 1/4,m\mp 1/4),\notag\\
(\eac,\edc)^-\simeq(n\pm 1/4,m\pm 1/4).
\label{res}
\end{align}
The first equation in Eq.\,(\ref{res}) states that in addition to Eq.\,(\ref{acdc}) there exists another condition, i.e. $\eac-\edc=m-1/2$, and the complete set of equations describing resistance maxima is
\begin{equation}
\eac + \edc \simeq n,\,\,\,\eac - \edc \simeq m-1/2.
\label{res2}
\end{equation}

We now look at the $r$ map near half-integer values, i.e. $(\eac,\edc)=(7/2,1/2)$ which is shown in Fig.\,\ref{image}(b). 
This resistivity map appears similar to the one near $(\eac,\edc)=(3,1)$ shown in Fig.\,\ref{image}(a) with one exception: maxima and minima are shifted further from the image center.
This difference is explained by the fact that MIRO extrema occur at $\eac=\eac^\pm\simeq n \mp \pac$,\citep{zudov:2004,studenikin:2005} where $\pac$ is considerably smaller $(\simeq 0.15$ in our case) [cf., vertical dashed lines] than a theoretical value of $1/4$.
As a result, the $\eac$ of the maxima are more precisely determined by the positions of the MIRO extrema observed in linear response. 
While one can easily modify Eq.\,(\ref{res}) to account for reduced phase of MIRO, we chose to leave it as it is to simplify our discussion and comparison with future theories. 
Fig.\,\ref{image}(a) and (b) also reveal existence of the saddle points in the resistivity map near $(\eac,\edc)=(n,m)$ and $(n+1/2,m+1/2)$, respectively.
We thus conclude that any\citep{note8} $(\eac,\edc)=(n/2,m/2)$ correspond to a saddle point of the resistivity.

\begin{figure}[t]
\includegraphics{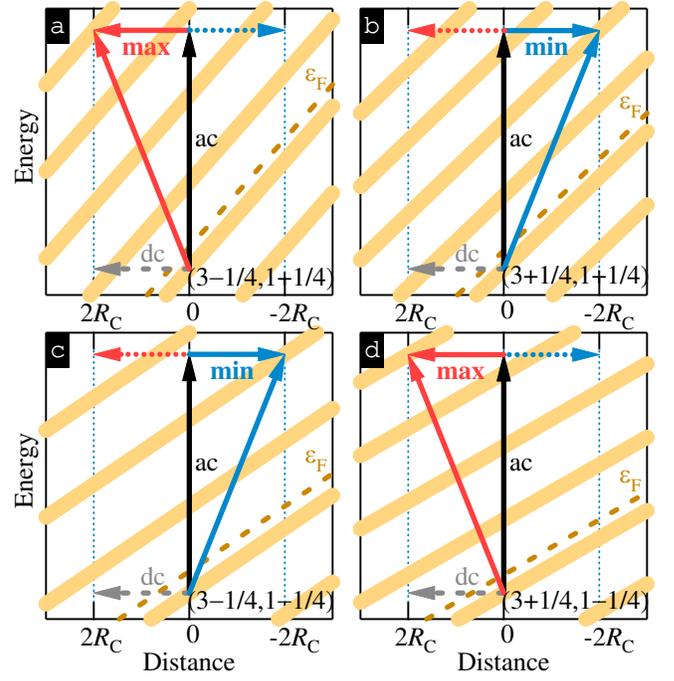}
\caption{
Thick lines represent LLs tilted by dc electric field.
Vertical (horizontal) arrows represent transitions due to microwave absorption (impurity scattering).
Inclined arrows show combined transitions. 
Conditions for the maxima [minima] are satisfied in (a) and (d) [(b) and (c)], respectively.
}
\vspace{-0.15 in}
\label{lls}
\end{figure}

Physically, conditions for the maxima given by Eq.\,(\ref{res2}) indicate that not only scattering in the direction parallel to the electric field needs to be maximized [first equation], but also that scattering in the opposite direction, i.e. antiparallel to the electric field, has to be minimized [second equation].
To illustrate this important point, we sketch LLs and relevant electron transitions in Fig.\,\ref{lls} as prescribed by Eqs.\,(\ref{res}), for $n=3$ and $m=1$. 
Panels (a) and (d), which correspond to the resistivity maxima at $(\eac,\edc)=(3\mp 1/4,1\pm/1/4)$, depict transitions to the left ending at the LL center (maximized) and transitions to the right terminating in the cyclotron gap (minimized). 
Similarly, panels (b) and (c) illustrate the conditions for the resistivity minima at $(\eac,\edc)=(3\pm 1/4,1\pm/1/4)$, with suppressed transitions to the left and enhanced transitions to the right.
Direct comparison of (a) and (c) [(b) and (d)] illustrates how the electric field detunes [tunes] transitions to the left while simultaneously tuning [detuning] transitions to the right, converting a maximum [minimum] to a minimum [maximum].

We can also understand oscillation suppression observed at $\eac=n+1/2$ in Fig.\,\ref{edc}(a) by taking into account impurity scattering events {\em not} accompanied by microwave absorption [cf.\,arrows marked by ``dc'' in Fig.\,\ref{lls}].
Let us consider two characteristic levels of dc excitation, namely $\edc=1/2$ and $\edc=1$.
At $\edc=1/2$, elastic transitions end in the cyclotron gap and the correction to the differential resistance is negative.\citep{zhang:2007a,vavilov:2007,lei:2007a}
Corresponding combined transitions (both parallel and antiparallel to $E$) terminate at the center of the LL, since both $\eac+\edc$ and $\eac-\edc$ are integers.
However, these transitions will not compensate each other because of the difference in electron distribution function at the final points.
Instead, transitions parallel to the electric field will win over the transitions antiparallel to the electric field and the correction to the differential resistance from both combined processes will be positive.
At $\edc=1$, the situation is reversed; elastic processes are enhanced and give positive correction while combined transitions give negative correction to the differential resistivities.
Thus, elastic and combined transitions tend to compensate each other in both cases, which result in strongly suppressed oscillation amplitude observed at half-integer $\eac$.

It would be interesting to study the interplay between combined and elastic scattering processes.
This can be achieved by tuning the microwave intensity.
According to our picture, at high enough microwave power, combined processes should dominate the response and oscillations at $\eac\simeq n+1/2$ will reappear.
These oscillations should be out of phase compared to the oscillations observed at $\eac\simeq n$ (or weak microwave intensities).
Since our microwave source was already running at full power, we were unable to test this prediction.

Finally, impurity scattering events without microwave absorption are expected to slightly modify Eq.\,(\ref{res}), by introducing additional shift along $\edc$ axis.
This shift is directed towards (away from) $\edc=n$ for the maxima (minima), respectively, and can be discerned in our data.
We therefore conclude that while combined transitions dominate the response, elastic scattering processes also play an important role.

In summary, we have studied nonlinear resistivity of a high-mobility 2DES under microwave illumination.
First, we have obtained complete sets of conditions for the differential resistivity maxima, minima, and saddle points.
Second, we have observed that the amplitude of dc-induced oscillations is strongly suppressed at half-integer values of $\eac$.
These observations indicate the importance of combined scattering against electric field and impurity scattering without microwave absorption, respectively, providing guidance for emergent theories.\citep{lei:2007b,auerbach:2007}
We expect that similar considerations should be applicable for other types of resistance oscillations, such as acoustic phonon resonances\citep{zudov:2001b} recently observed in non-linear resistivity.\citep{zhang:2008}

We gratefully acknowledge discussions with M. Khodas and M. Vavilov.
This work was supported by NSF Grant No. DMR-0548014.

\begin{thebibliography}{19}
\bibitem[{not()}]{note8}
\bibinfo{note}{Except $n=m$, which is not a saddle point.}
\end{thebibliography}

\end{document}